\documentclass{PoS}

\usepackage{axodraw}

\PoS{PoS(LAT2005)058}

\newcommand{\be}{\begin{equation}}
\newcommand{\ee}{\end{equation}}
\newcommand{\ba}{\begin{eqnarray}}
\newcommand{\ea}{\end{eqnarray}}
\newcommand{\dsp}{\displaystyle}

\newcommand{\blue}{\color{blue}}
\newcommand{\green}{\color{green}}
\newcommand{\red}{\color{red}}

\title{Two Loop Partially Quenched and Finite Volume Chiral Perturbation Theory Results }

\ShortTitle{Two Loop Partially Quenched and Finite Volume Chiral Perturbation Theory Results }

\author{\speaker{Johan Bijnens}
        \\
        Department of Theoretical Physics, Lund University,
        S\"olvegatan 14A, SE 22362 Lund, Sweden\\
        E-mail: \email{bijnens@thep.lu.se}}

\author{Niclas Danielsson\\
        Department of Theoretical Physics, Lund University,
        S\"olvegatan 14A, SE 22362 Lund, Sweden\\
and\\Division of Mathematical Physics, LTH,
Lund University, Box 118, S 221 00 Lund, Sweden
        E-mail: \email{niclas.danielsson@matfys.lth.se}}

\author{Karim Ghorbani\\
        Department of Theoretical Physics, Lund University,
        S\"olvegatan 14A, SE 22362 Lund, Sweden\\
        E-mail: \email{karim@thep.lu.se}}

\author{Timo L\"ahde\\
        Department of Theoretical Physics, Lund University,
        S\"olvegatan 14A, SE 22362 Lund, Sweden\\
        E-mail: \email{talahde@thep.lu.se}}

\abstract{
This talk presents some results relevant for lattice QCD at higher order
in ChPT. First we discuss the finite volume corrections at two loops
for the quark condensate as well as a L\"uscherlike finite volume formula
for it. The latter allows for an alternative determination of meson sigma
terms. The second set of results presented here are
the calculations at two loops in partially quenched chiral perturbation theory
of masses and decay constants of the charged mesons.
We present results
for all relevant quark mass combinations for the cases with two and three
sea quarks.}

\FullConference{ 
XXIIIrd International Symposium on Lattice Field Theory\\
                 25-30 July 2005\\
                 Trinity College, Dublin, Ireland}

\renewcommand{\logo}
{\raisebox{-5mm}
{\parbox{\textwidth}{\begin{flushright}LU TP 05-34\\
hep-lat/0509042\\
September 2005
\end{flushright}
\includegraphics{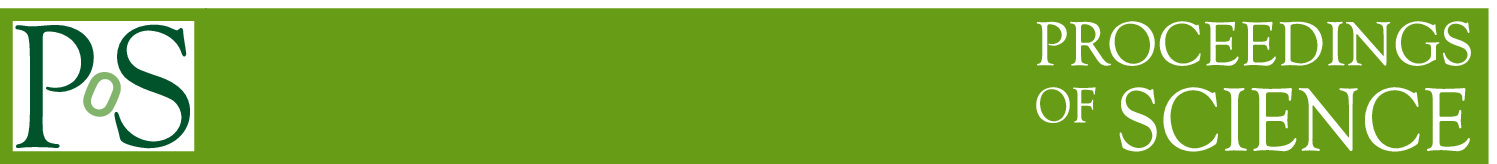}}
}}
\FullConference{{\rm To be published in the proceedings of:\\} 
XXIIIrd International Symposium on Lattice Field Theory\\
                 25-30 July 2005\\
                 Trinity College, Dublin, Ireland}

\begin{document}

\section{Introduction}

In this talk we present some work useful for the extrapolation of lattice QCD
numerical results to the physical volume and quark masses.
At low energies, there exists an effective field theory approximation to
QCD known as Chiral Perturbation Theory (ChPT) \cite{GL}. We can thus use
this theory to fit to lattice QCD numerical results and then extrapolate
to infinite volume and the physical quark masses. ChPT is based on
the correct inclusion of the nonanalytic structure in all amplitudes due to
the Goldstone bosons of spontaneous chiral
symmetry breaking and all relations that
follow from the chiral Ward identities. This is done in a systematic expansion
in momenta, energies and quark masses.
There exists an enormous body of work in ChPT. Meson masses, decay
constants and vacuum expectation values at infinite volume
are known to two-loop order, see \cite{ABT3,ABT1,ABT2} and references therein.
In this talk we present the extension of these results in two directions,
finite volume for vacuum expectation values and partial quenching for the
masses and decay constants.

\section{$\langle\overline q q\rangle$ at Finite Volume}

The work presented in this section will be published in \cite{BG}. 
At finite volume, particles can propagate around the world. Taking this into
account in ChPT was first done in \cite{GLfinite}.
Many more quantities have since been calculated at the one-loop level.
Here we would like to
address this problem at the two-loop level. A major problem is to evaluate the
necessary loop integrals at two-loop order which has not been done
so far\footnote{Some partial results were presented by C. Haefeli at
this conference.}. An alternative approach to finite volume was pioneered
by L\"uscher \cite{Luescher}. Here the leading effect  of a finite volume
is taken into account by an integral over another amplitude.
An overview of earlier finite volume work can be found in \cite{Colangelo}.

\begin{figure}
\begin{minipage}{0.4 \textwidth}
\includegraphics[width=\textwidth]{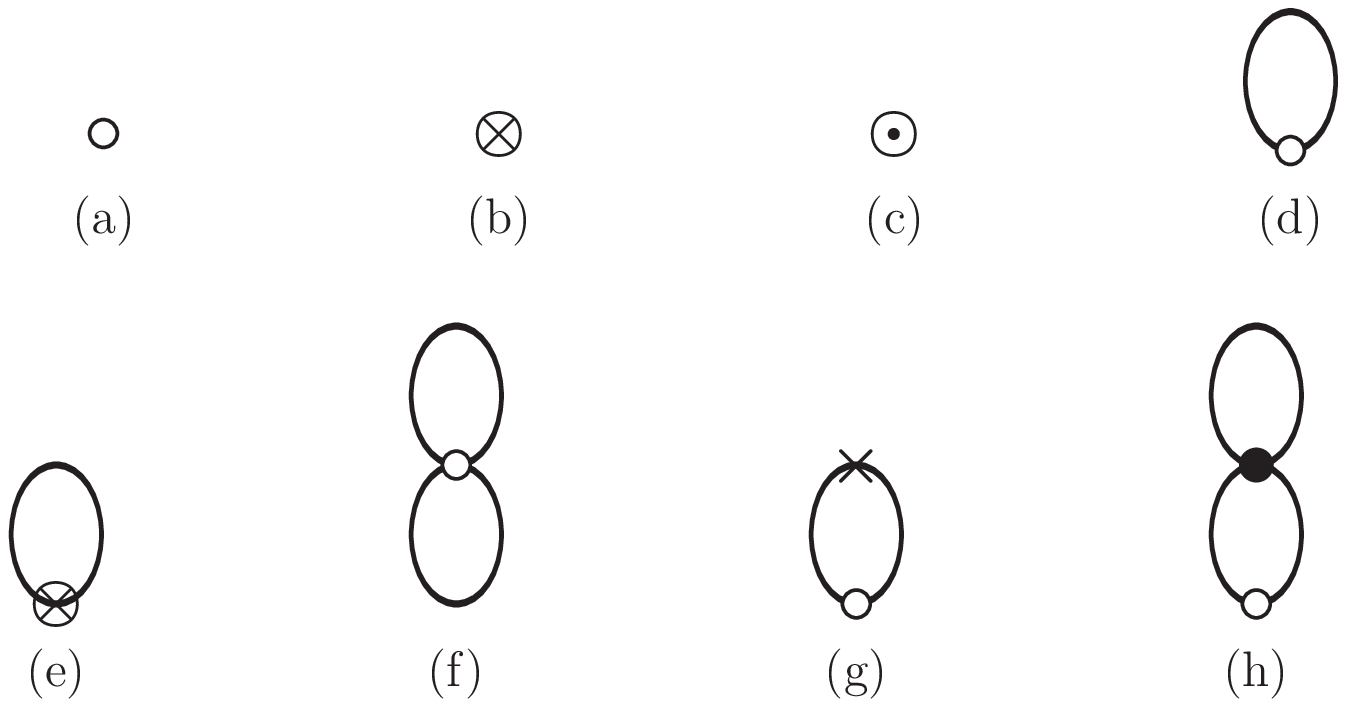}
\caption{The diagrams needed for $\langle \overline q q\rangle$.
{Lines:} meson propagators;
{Insertion of $\overline qq$:} $\circ$  ($p^2$), 
$\otimes$ ($p^4$), $\odot$ ($p^6$);
{Vertex:} $\bullet$ ($p^2$), $\times$ ($p^4$).}
\label{fig1}
\end{minipage}
\begin{minipage}{0.58 \textwidth}
\includegraphics[angle=-90,width=\textwidth]{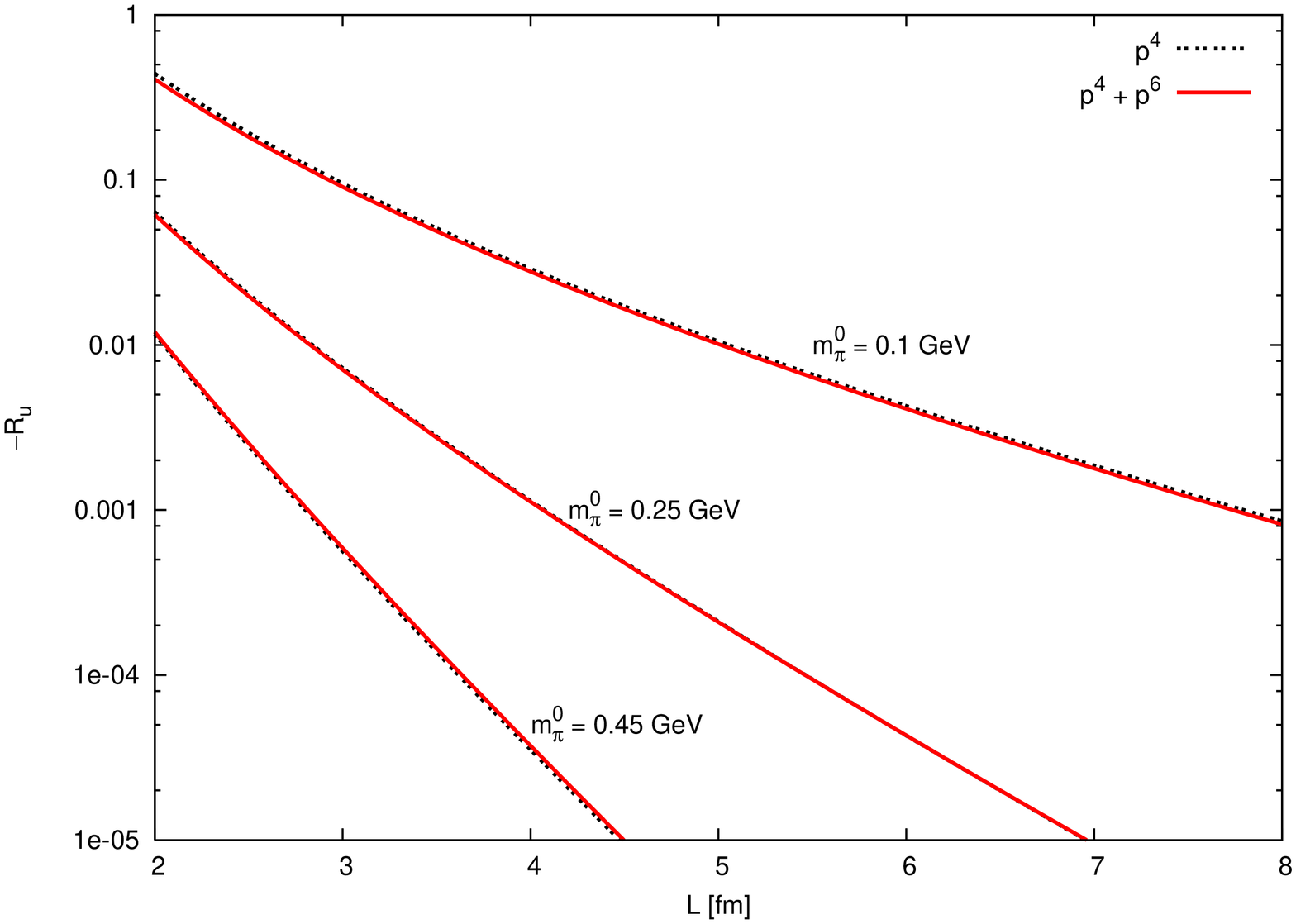}
\caption{The finite volume corrections $R_u$ to $\langle\overline uu\rangle$.}
\label{fig2}
\end{minipage}
\end{figure}

We would like to check how well L\"uscher's method works when the finite
volume quantities are fully calculated. At one loop in ChPT the full results for
many quantities can be derived using L\"uscher's method. We therefore want to
study a quantity at two-loop level. 

There are a few quantities where the
two-loop calculation does not involve irreducible two-loop integrals. One such
example is precisely the value of $\langle \overline q q\rangle$ as can be seen
from the diagrams shown in Fig.~\ref{fig1}. $\langle \overline q q\rangle$
at two-loops at infinite volume is known \cite{ABT1}. We have redone this
calculation checking explicitly that the ${\cal O}(d-4)$ parts of the
integrals do not contribute to the final results. The known one-loop integrals
at finite volume are therefore sufficient. Numerical results
for
$
R_q = 
\left({\langle\overline qq\rangle_V-\langle\overline qq\rangle_\infty}
\right)/
{\langle\overline qq\rangle_\infty}
$
are shown in Fig.~\ref{fig2} for the case of the up quark condensate.
We have here expressed the masses appearing in the integrals as the lowest
order masses, leading to very small ${\cal O}(p^6)$ corrections.

\begin{figure}
\begin{minipage}{0.35\textwidth}
  \includegraphics[clip,width=\textwidth]{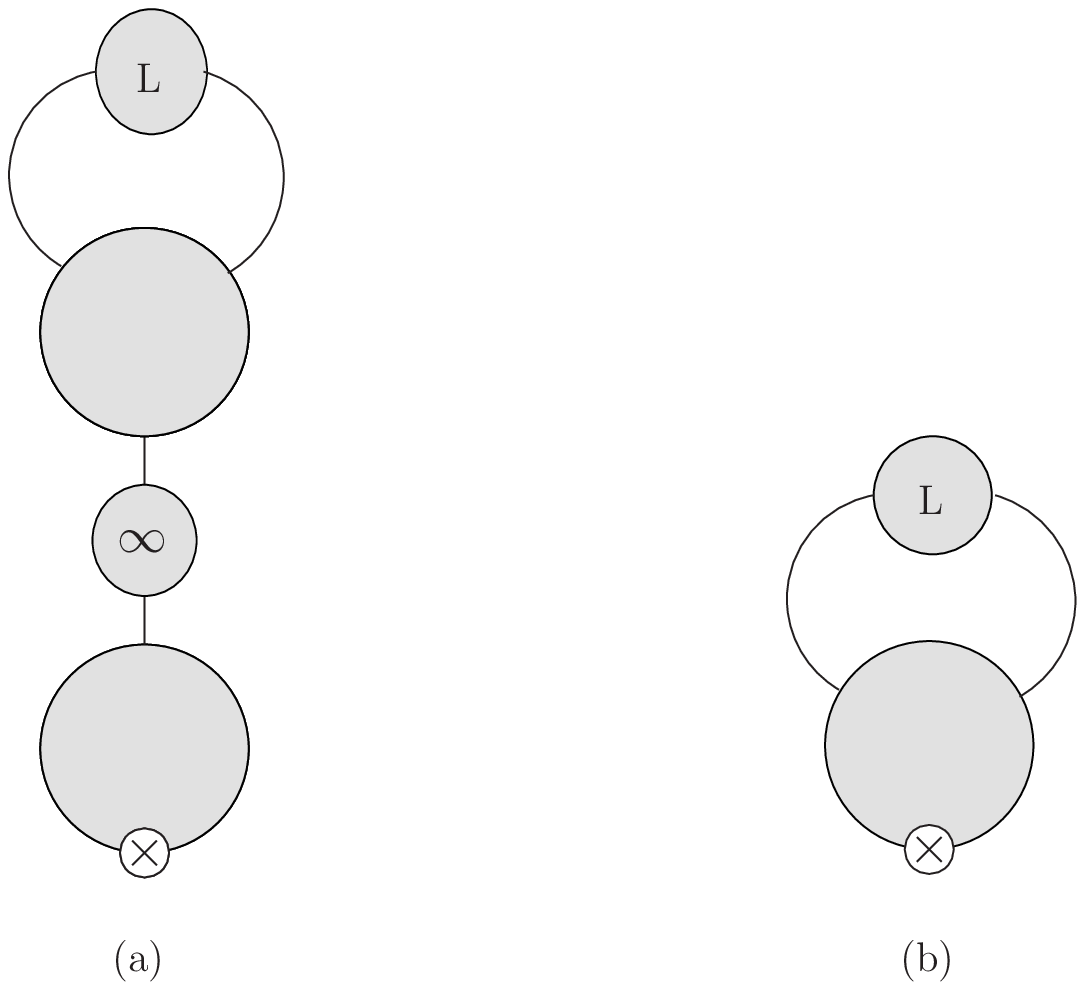}
  \caption{The two classes of contributions to a L\"uscher formula for a vacuum
 expectation value. The label $L$ or $\infty$ means a finite volume or infinite
 volume propagator.}
  \label{fig3}
\end{minipage}
~
\begin{minipage}{0.59\textwidth}
\begin{center}
\unitlength=0.5pt
\begin{picture}(100,130)
\SetScale{0.5}
\SetWidth{3.}
\Text(50,100)[]{Mesons}
\SetColor{Red}
\Line(0,10)(50,10)
\Vertex(50,10){6}
\CArc(50,40)(30,0,180)
\CArc(50,40)(30,180,360)
\Line(50,10)(100,10)
\end{picture}
~{\Large $=$}~
\unitlength=0.5pt
\begin{picture}(100,130)
\SetScale{0.5}
\SetWidth{2.}
\Text(50,126)[]{Quark Flow}
\Text(50,95)[]{\blue Valence}
\SetColor{Blue}
\ArrowLine(35,10)(0,10)
\ArrowArc(50,40)(35,-60,240)
\ArrowLine(100,10)(65,10)
\ArrowLine(0,2)(45,2)
\Line(45,2)(45,13)
\ArrowArcn(50,40)(28,260,-80)
\Line(55,13)(55,2)
\ArrowLine(55,2)(100,2)
\end{picture}
~{\Large $+$}~
\unitlength=0.5pt
\begin{picture}(100,130)
\SetScale{0.5}
\SetWidth{2.}
\Text(50,126)[]{Quark Flow}
\Text(50,95)[]{\blue Valence}
\SetColor{Blue}
\ArrowLine(35,10)(0,10)
\ArrowArc(50,40)(35,-60,240)
\ArrowLine(100,10)(65,10)
\ArrowLine(0,2)(50,2)
\ArrowLine(50,2)(100,2)
\ArrowArcn(50,40)(28,270,-90)
\end{picture}
~{\Large $+$}~\\[0.5cm]
\unitlength=0.5pt
\begin{picture}(100,130)
\SetScale{0.5}
\SetWidth{2.}
\Text(50,126)[]{Quark Flow}
\Text(50,95)[]{\green Sea}
\SetColor{Blue}
\ArrowLine(35,10)(0,10)
\ArrowArc(50,40)(35,-60,240)
\ArrowLine(100,10)(65,10)
\ArrowLine(0,2)(50,2)
\ArrowLine(50,2)(100,2)
\SetColor{Green}
\ArrowArcn(50,40)(28,270,-90)
\end{picture}
~{\Large$+$}~
\unitlength=0.5pt
\begin{picture}(100,130)
\SetScale{0.5}
\SetWidth{2.}
\Text(50,126)[]{Quark Flow}
\Text(50,95)[]{\red Ghost}
\SetColor{Blue}
\ArrowLine(35,10)(0,10)
\ArrowArc(50,40)(35,-60,240)
\ArrowLine(100,10)(65,10)
\ArrowLine(0,2)(50,2)
\ArrowLine(50,2)(100,2)
\SetColor{Red}
\ArrowArcn(50,40)(28,270,-90)
\end{picture}
\end{center}
\caption{A schematic representation of how a meson loop contribution has
quark flows with valence, sea and ghost quark closed loops present.}
\label{fig4}
\end{minipage}
\end{figure}

The analogue of L\"uscher's mass formula is derived in
\cite{BG}. The two types of contributions are shown in Fig.~\ref{fig3}.
Those of Fig.~\ref{fig3}a do not contribute in ChPT to
parity-even operators like  $O=\overline qq$.
Keeping also the non-leading contributions of the same propagator,
see \cite{Colangelo}, we obtain
\ba
\dsp
{\langle O \rangle_V-\langle O \rangle_\infty} &=&
-\sum_{\vec n\ne \vec o} \frac{1}{16\pi^2}
\int_0^\infty \frac{dq^2 q^2}{\sqrt{m_0^2+q^2}} e^{-\sqrt{\vec n^2
(m_0^2+q^2)L^2}} \langle\phi| O | \phi\rangle
\nonumber\\
& =& -{ \langle\phi| O | \phi\rangle}
\left(\sum_{k=1,\infty}\frac{m(k)}{16\pi^2} \frac{m_0^2}{\sqrt{\zeta(k)}} 
K_1(\zeta(k))\right)\,.
\ea
with $\zeta(k)=\sqrt{k}\,m_0 L$ for the contribution of a neutral scalar $\phi$
to the finite volume dependence of $\langle O\rangle$. $m(k)$ is the number of
times $k=\sum_{i=1,3}n_i^2$ appears in the sum over $\vec n$.

Note that in this case there is a universal factor which contains the
matrix element $\langle\phi|O|\phi\rangle$ rather than an integral as in the
finite volume mass formula. This also means that the finite volume dependence
can be used to obtain this matrix element. Unfortunately, the extremely
small ${\cal O}(p^6)$ corrections allow no meaningful test of
the L\"uscher formula versus the full ${\cal O}(p^6)$ calculation.

\section{Partially Quenched ChPT at Two Loops}

The extension of ChPT to the quenched and partially quenched case has been
done at one loop by Morel, 
Bernard, Golterman and Sharpe \cite{MorelBG1SharpeA,BG2}
and discussed in detail in \cite{Sharpe1Sharpe2}. We use here the version
without the super-singlet $\Phi_0$.
Basically one adds bosonic ghost quarks to cancel the effect of valence
loops and
a separate set of sea quarks to the Lagrangian of QCD, shown schematically in
Fig.~\ref{fig4}.
 As a consequence
the Lagrangian gets (approximately, see \cite{Sharpe1Sharpe2}) 
a graded symmetry
 $SU(n_v+n_s|n_v)\times SU(n_v+n_s|n_v)$, 
with $n_v$ the number of valence and $n_s$ the number of sea quarks,
rather than the usual chiral symmetry.
We thus expect the low-energy effective theory to be Chiral Perturbation Theory
with this symmetry group instead. Most of the classification work and the
infinity structure at two loops \cite{BCE1} can be carried over to this
sector by simply replacing traces by super-traces. Exceptions to this are
baryons and the Cayley-Hamilton relations.

One important note is that since QCD is a continuous limit of partially
quenched QCD (PQQCD) the low-energy constants (LECs) from ChPT
can be derived directly from
those of PQChPT, e.g.
\be
 L_1^r = L_0^{r(3pq)}/2 + L_1^{r(3pq)}\,.
\ee
The superscript $(3pq)$ means the constants relevant for 3 sea quark flavour
PQChPT.

The calculations published so far are: The mass at two loops in PQChPT
for the case of equal valence and equal sea-quark masses for three 
sea
flavours \cite{BDL1}, the decay constants for three sea flavours \cite{BL1}
and the masses and decay constants for two sea flavours \cite{BL2}.
The remaining mass combinations for the masses are in the process of being
written up \cite{BDL2}. That article will also contain a longer introduction to
PQChPT at two-loop order.

The expressions are extremely long compared to those of the normal
case \cite{ABT3}. 
The main reasons are: First, 
there are many more masses even compared to the isospin
breaking case \cite{ABT2}. This allows for many more mass combinations to
show up. Second and most important, the neutral or diagonal sector propagators
contain both single and double poles with complicated coefficients.
These coefficients are ratios of differences of lowest order meson masses
to various powers. These rational functions of quark masses satisfy many
nontrivial relations which allowed to compress the formulas by
significantly more than an order of magnitude, and bring them to a
publishable size, but they still leave an added layer of complexity.

The double poles in the propagators appear because
PQQCD is not a well defined field theory but
only exists as a thermodynamical model. This leads to the well-known
quenched logarithms~\cite{MorelBG1SharpeA}.

In the remainder we present some numerical results at two-loop order for
masses and decay constants. We present them as a function of quark masses, 
$m_i$,
via the lowest
order meson masses
\be
\chi_i = m_M^{2(0)} \equiv 2 B_0 m_i\,.
\ee
 The border of validity for ChPT is for the 
$\chi_i$ of order $0.25\sim0.30$~GeV$^2$. 

\begin{figure}
\begin{minipage}{0.46\textwidth}
\includegraphics[width=\textwidth]{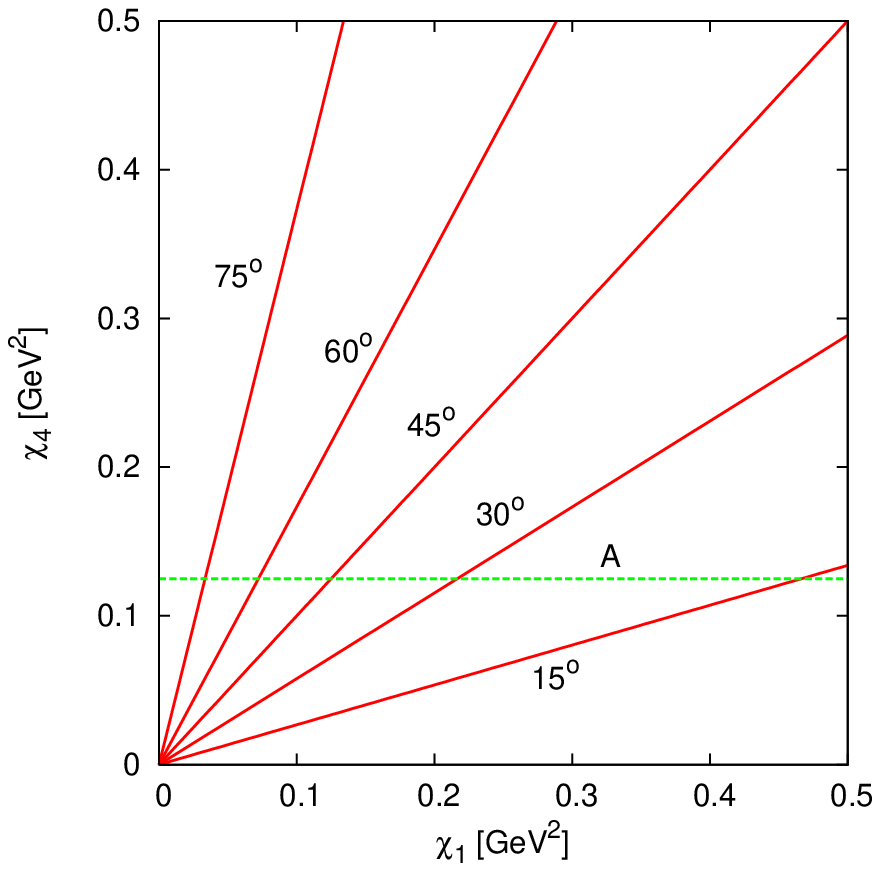}
\caption{The lines in the sea/valence quark mass along which results are shown
in the next figures.}
\label{fig5}
\end{minipage}
~
\begin{minipage}{0.46\textwidth}
\includegraphics[width=\textwidth]{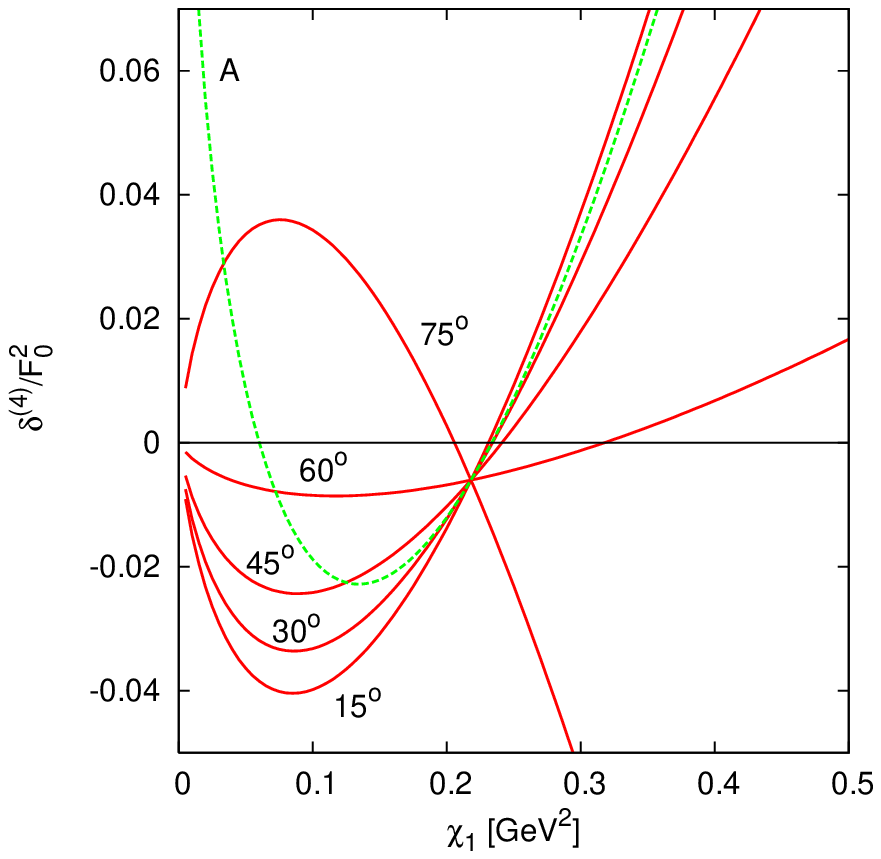}
\caption{The relative correction to the masses at order $p^4$
for the 1+1 case \cite{BDL1}.}
\label{fig6}
\end{minipage}
\end{figure}
\begin{figure}[b]
\begin{minipage}{0.46\textwidth}
\includegraphics[width=\textwidth]{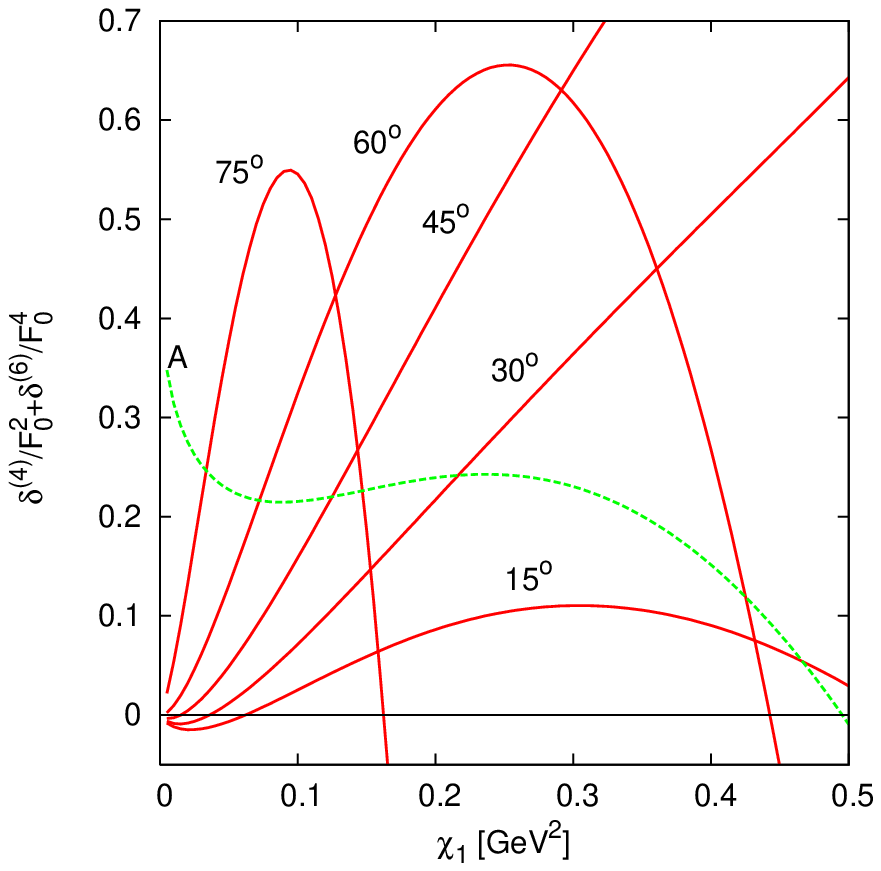}
\caption{The relative correction to the masses at order $p^6$
for the 1+1 case \cite{BDL1}.}
\label{fig7}
\end{minipage}
~
\begin{minipage}{0.46\textwidth}
\includegraphics[width=\textwidth]{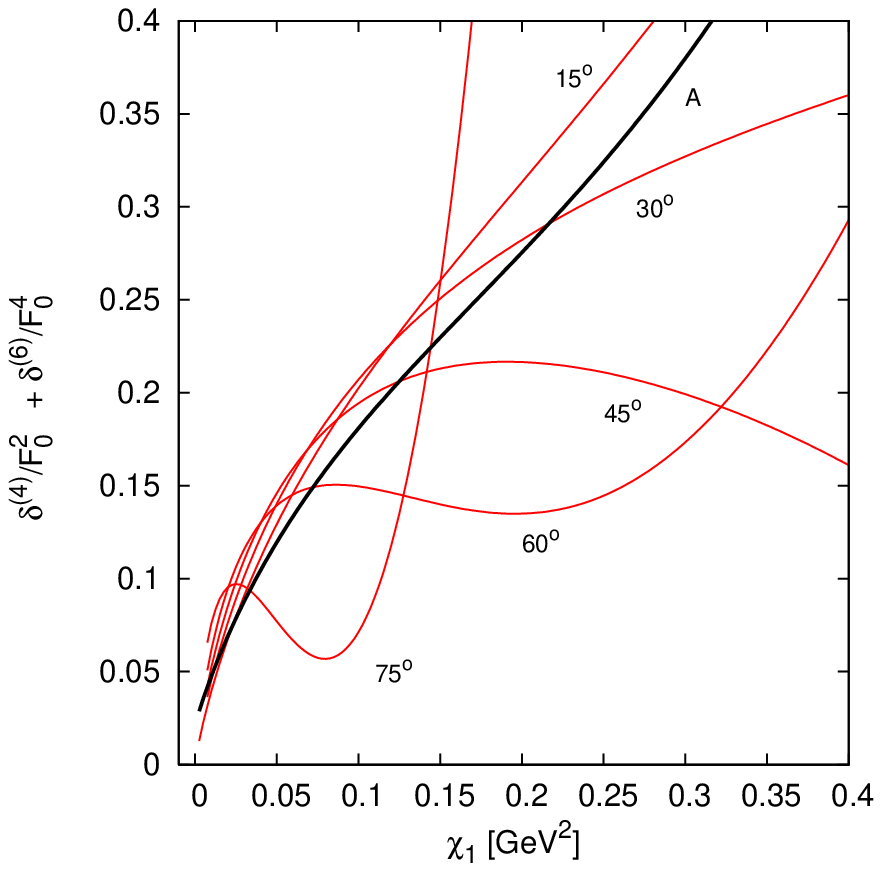}
\caption{The relative correction to the decay constants at order $p^6$
for the 1+1 case \cite{BL1}.}
\label{fig8}
\end{minipage}
\end{figure}

One can of course show any amount of figures for the various mass cases
for decays and masses. We restrict to a few simple cases. First 
the 1+1 case, i.e. valence masses equal:$\chi_1=\chi_2=\chi_3$ and 
sea masses equal: $\chi_4=\chi_5=\chi_6$.
We show results along the lines $\chi_4=\tan\theta\,\chi_1$ and $\chi_4$
constant, shown in
Fig.~\ref{fig5} in the valence/sea quark plane.
The result for the mass at one-loop order is shown in Fig.~\ref{fig6}.
Notice the presence of the quenched chiral logarithm along the line labeled~A.
The same result at two-loop order is shown in Fig.~\ref{fig7}. The corrections
are much larger than at one-loop order. This is also the case for normal ChPT
\cite{ABT2}. There are choices of the LECs possible where this is not
so \cite{ABT1,BD}.
We also show results for the 1+1 case of the decay constant \cite{BL1}.

Plots for the other mass cases can be found in the published papers.
The analytical form of the expressions can be downloaded from \cite{webpage}
and numerical programs are available from the authors.

In conclusion for this part, we have obtained the two-loop formulae for
charged meson masses and decay constants in PQChPT and shown that the
corrections can be sizable. We hope that this work will be useful for the
numerical lattice QCD community.

\section*{Acknowledgments}

This work is supported by the European Union TMR network,
Contract No. 
HPRN-CT-2002-00311  (EURIDICE) and by 
the European Community-Research Infrastructure
Activity Contract No. RII3-CT-2004-506078 (HadronPhysics).
TL acknowledges a Mikael 
Bj\"ornberg memorial foundation travel grant
and KG an Iranian
government fellowship.

\end{document}